\documentclass[review,12pt]{elsarticle}
\usepackage{lineno,hyperref}
\modulolinenumbers[5]

\journal{Journal of \LaTeX\ Templates}

\usepackage{amsmath}
\usepackage{graphicx}

\title{Amplification of Hypersound in Graphene with degenerate energy dispersion}

\author[rvt]{K. A. Dompreh\corref{cor1}\fnref{fn1}}
\author[focal]{N. G. Mensah}
\author[rvt]{S. Y. Mensah}
\address[rvt]{Department of Physics, College of Agriculture and Natural Sciences, U.C.C, Ghana.}
\address[focal]{Department of Mathematics, College of Agriculture and Natural Sciences, U.C.C, Ghana}
\cortext[cor1]{Corresponding author\\ Email: kwadwo.dompreh@ucc.edu.gh}
\ead[url]{kwadwo.dompreh@ucc.edu.gh}

\date{}
\begin{document} 
\begin{abstract}
Hypersound amplification/absorption of acoustic phonons  in Graphene with  
degenerate energy dispersion $\varepsilon(p)$ near the Fermi level was theoretically studied. 
For $k_{\beta}T << 1$ and $ql >> 1$, the dependence of the absorption coefficient $\Gamma/\Gamma_0$ on ${V_D\over V_s}$ was studied where the  
results satisfied the Cerenkov effect. That is when ${V_D\over V_s} > 1$, an amplification was
obtained but for  ${V_D\over V_s} < 1$, an absorption was obtained which could lead to Acoustoelectric Effect (AE) in Graphene.
A linear  dependence of the $\Gamma/\Gamma_0$ on $\omega_q$ was observed where the result  
obtianed qualitatively agreed with  an experimentally observed  acoustoelectric current in Graphene
via the  Weinrich relation. It is interesting to note  from this study that, frequencies above $10THz$
can be attained for $V_D = 1.1ms^{-1}$. This study permit the use of Graphene as hypersound phonon laser (SASER). 

\end{abstract}

\maketitle
\section*{Introduction}
Graphene, a member of the Carbon allotropes has  exceptional properties 
for future nanoelectronics ~\cite{1,2,3,4}. It is an ideal two-dimensional electron gas ($2DEG$) 
system made up of one layer of Carbon atom  having a high electron mobility ($\mu$)
at room temperature with high mechanical and thermodynamic stability ~\cite{5}.
Several unusual phenomena  such as half-integer quantum Hall effect~\cite{6}, non-zero 
Berry's phase~\cite{7}, and minimum conductivity~\cite{8}  have been 
observed experimentally in Graphene. The most interesting property of Graphene is its  linear 
energy dispersion $E =\pm \hbar V_F \vert k \vert$ (the Fermi velocity $V_F \approx 10^8ms^{-1}$) at the Fermi level with 
low-energy excitation. This makes graphenes 
applicable  in advance electronics and optoelectronic devices such as sub-terahertz  Field-effect 
transistors~\cite{9}, infrared transparent electrodes~\cite{10} and $THz$ plasmonic deives~\cite{11}. 
Currently, among the various studies on Graphene  attracting much attention  is the generation and detection 
of hypersound  amplification or absorption of acoustic phonons~\cite{12}. 
It is known that, when an acoustic phonon passes through a semiconductor, it 
may interact with various elemental excitations which may lead to amplifcation or absorption  
of the phonons.  The idea of acoustic wave amplification in bulk material was theoretically 
predicted by Tolpygo (1956), Uritskii~\cite{13}, and  Weinreich~\cite{14} and in N-Ge by Pomerantz~\cite{15}. 
Hypersound generation  in bulk ~\cite{16} and 
low-dimensional materials such as Superlattices~\cite{17,18,19,20}, Cylindrical Quantum 
Wire~\cite{21}, Quantum Wells~\cite{22} and Graphenes Nanoribbons (GNR) ~\cite{23} have been 
studied. Akin to Cerenkov acoustic-phonon emission, when the drift velocity of electrons  $V_D$ 
exceeds the sound velocity ($V_s$) of the host material~\cite{24} lead to amplification of the 
acoustic-phonons or when $V_D < V_s$ causes absorption. This has been ultilised experimentally to confirm 
the breakdown of quantum Hall effect~\cite{25}, the generation of coherent phonon-polariton radiation~\cite{26}, 
and large acoustic gain in coherent phonon oscillators in semiconductors~\cite{27}. 
Furthermore, the emission and absorption  of acoustic-phonons is used to provide detailed information on the 
excitation and relaxation mechanisms in semiconductors via deformation potential, where  the effect of  interactions 
can be used to determine the physical properties of the material. In particular, acoustic-phonons 
providing terahertz ($10^{12}Hz$) hypersonic sources can lead to the attainment of phonon laser or SASER~\cite{28,29} 
in graphene via Cerenkov effect which is an intense field of research. Following the works of Nunes and Fonseca~\cite{32}, Zhao et. al~\cite{33} proposed 
the possibility of attaining Cerenkov acoustic-phonon emission in Graphene whilst Insepov et. al~\cite{31}, performed experimentally the surface acoustic wave Amplification by D.C voltage supply in 
Graphene. In this paper, unlike ~\cite{32}, where when $V_D = 0$, expression goes to zero, in otherwords there
is no absorption. Here, we have a general expresion which acounts for absorption when $V_D = 0$ or $V_D < V_s$. This 
leads to the observation of acoustoelectric effect. This has been verified experimentally in Graphene~\cite{37}.
The  motivation for this work is to provide the theoretical framework that can lead to the attainment 
of SASER in Graphene, for use  as a phonon spectrometer,  for generation 
of high-frequency electric oscillation, and as a non-destructive testing of microstructure and acoustic 
scanning system.   The paper is organised as follows: In theory section, 
the theory underlying the amplification (Absorption) of acoustic-phonon via Cerenkov effect is presented.
In the  numerical analysis section,  the final equation is analysed  and presented in a  graphical form.   
Lastly, the  conclusion is presented in section $4$.

\section*{Theory}
We will proceed following the works of ~\cite{32}, here the acoustic wave will be consisdered as phonons of frequency ($\omega_q$) in 
the short-wave region  $ql >> 1$ ($q$  is the acoustic wave number, $l$ is the electron mean free path). The kinetic equation for the acoustic 
phonon population $N_{\vec{q}}(t)$ in the graphene sheet is given by

\begin{eqnarray}
\frac{\partial N_{\vec{q}}}{\partial t} =\frac{2\pi}{\hbar}g_s g_v\sum_{k,k^{\prime}}\vert{C_{\vec{q}}}\vert^2 \delta_{k,k^{\prime}+{\vec{q}}}
\{[N_{\vec{q}}(t) + 1]f_{\vec{k}}(1-f_{\vec{k}^\prime})
\delta(\varepsilon_{\vec{k}^\prime} - \varepsilon_{\vec{k}} +\hbar\omega_{\vec{q}})\nonumber\\
- N_{\vec{q}}(t) f_{\vec{k}^\prime}(1-f_{\vec{k}})\delta(\varepsilon_{\vec{k}^\prime} - \varepsilon_{\vec{k}} - \hbar\omega_{\vec{q}})\}\label{Eq_1}
\end{eqnarray}
where $g_s = g_v = 2$ account the for spin and valley degeneracies respectively, $N_{\vec{q}}(t)$ represent the number of phonons with a wave vector $\vec{q}$ at time $t$. 
The factor $N_{\vec{q}} + 1$ accounts for the 
presence of $N_{\vec{q}}$ phonons in the system when the additional phonon is emitted. The $f_{\vec{k}}(1-f_{\vec{k}})$ represent the probability that 
the initial $\vec{k}$ state is occupied and the final electron state $\vec{k}^\prime$ is empty whilst the factor $ N_{\vec{q}} f_{\vec{k}^\prime}(1-f_{\vec{k}})$
is that of the boson and fermion statistics. The unperturbed electron distribution function is given by the shifted Fermi-Dirac
function as 
\begin{equation}
f_{\vec{p}} = [exp(-\beta(\varepsilon(\vec{p}- m v_D)-\chi))]^{-1}\label{Eq_2}
\end{equation}
where $f_{\vec{p}}$ is the Fermi-Dirac equilibrium function, with $\chi$ being the chemical potential, $\vec{p}$ is momentum 
of the electron, $\beta = 1/kT$, $k$ is the Boltzmann constant and $V_D$ is the net drift velocity relative to the ion 
lattice site. In Eqn ($1$), the summation over $k$ and $k^{\prime}$ can be transformed into integrals by the prescription
$$\sum_{k,k^\prime}\rightarrow \frac{A^2}{(2\pi)^4}\int d^2 k d^2k^{\prime}$$
where $A$ is the area of the sample, and assuming that $N_q(t) >> 1$ yields
\begin{equation}
\frac{\partial N_{\vec{q}}}{\partial t} = \Gamma_{\vec{q}}N_{\vec{q}}\label{Eq_3}
\end{equation}
where 
\begin{eqnarray}
\Gamma_{\vec{q}} = \frac{A\vert {\Lambda}\vert^2\hbar q}{(2\pi)^3 \hbar V_F\rho V_s}\int_0^\infty{kdk}\int_0^{\infty}{k^\prime dk^\prime}
\int_0^{2\pi}{d\phi}\int_0^{2\pi}{d\theta}\{[f(k)-f(k^\prime)]\nonumber \\
\delta(k-k^\prime-\frac{1}{\hbar V_F}(\hbar\omega_q - V_D\cdot\hbar\vec{q}))\} \label{Eq_4}
\end{eqnarray}
with  $ k^\prime = k - \frac{1}{\hbar V_F}(\hbar\omega_q - V_D\cdot \hbar\vec{q})$. ${\Lambda}$ is the deformation potential constant,
and $\rho$ is the density of the graphene sheet. At low temperature $k_B T << 1$, the distribution function become
$f(k) = exp(-\beta(\varepsilon(k)))$. Eqn($4$) can be expressed as 
\begin{eqnarray}
\Gamma_{\vec{q}} = \frac{A\vert {\Lambda}\vert^2\hbar q}{(2\pi) \hbar V_F\rho V_s}\int_0^\infty{kdk}
(k -\frac{1}{\hbar V_F}(\hbar\omega_q - V_D\cdot\hbar\vec{q}))[exp(-\beta \hbar V_F k) \nonumber \\
- exp(-\beta\hbar V_F(k - \frac{1}{\hbar V_F}(\hbar\omega_q - V_D\cdot\hbar\vec{q})))]\label{Eq_5}
\end{eqnarray}
Using standard intergrals, Eqn($5$) can be expressed finally 

as 
\begin{equation}
\Gamma = \Gamma_0 \{2 -\beta\hbar\omega_q(1-\frac{V_D}{V_s})\}[1- exp(-\beta\hbar\omega_q(1-\frac{V_D}{V_s}))]
\end{equation}
where 
\begin{equation}
\Gamma_0 = \frac{A\vert {\Lambda}\vert^2 q}{(2\pi){\beta}^3{\hbar}^3{ V_F}^4\rho V_s}
\end{equation}

\section*{Numerical Analysis}
The Eqn ($6$) is analysed numerically for a normalized graph of $\Gamma\over\Gamma_0$ against  $V_D\over V_s$ 
and $\omega_q$.  The following parameters were used $\Lambda = 9 eV$, $T = 10K$, $V_s = 2.1\times 
10^6cms^{-1}$ and $\vec{q} = 10^5 cm^{-1}$.
In Figure $1$, the graph for the  dependence of $\Gamma\over\Gamma_0$  on  
$\omega_q$ is plotted. The graph was obtained at ${V_D\over V_s} < 1$. 
The insert shows an experimentally  obtained graph of an acoustoelectric current  
for gate-controlled Graphene~\cite{37}. The hypersound absorption
graph qualitively agreed with the experimentally obtained graph via the Weinriech relation~\cite{36}.  
In Figure $2a$, the  dependence of $\Gamma\over\Gamma_0$  on  $V_D\over V_s$ is  
analysed. From the graph,  when  ${V_D\over V_s} > 1$, gives rise to an amplification 
 as also  indicated in the work of Nunes and Fonseca ~\cite{32}, however, when 
${V_D\over V_s} < 1$, an absorption of acoustic phonons was observed. To enhanced the observed Amplification (Absorption),
a $3D$ graph was plotted for frequencies $\omega_q = 0.2$, $0.4$, and  $1THz$ (see Figure $2b$, $3$ and $4$).
In Figure $2b$,  the maximum amplification was obtained at ${\Gamma\over\Gamma_0} = -0.16$ at 
$\omega_q = 2THz$ for $V_D = 1.1V_s$.  For figure $3(a)$,  at $V_D = 1.1 V_s$, ${\Gamma\over\Gamma_0} = -0.34$ 
whislt in figure $3(b)$,  for $V_D = 1.1 V_s$, ${\Gamma\over\Gamma_0} = -0.08$ was obtained. 
\begin{figure}
\begin{centering}
\includegraphics[width = 9.0cm]{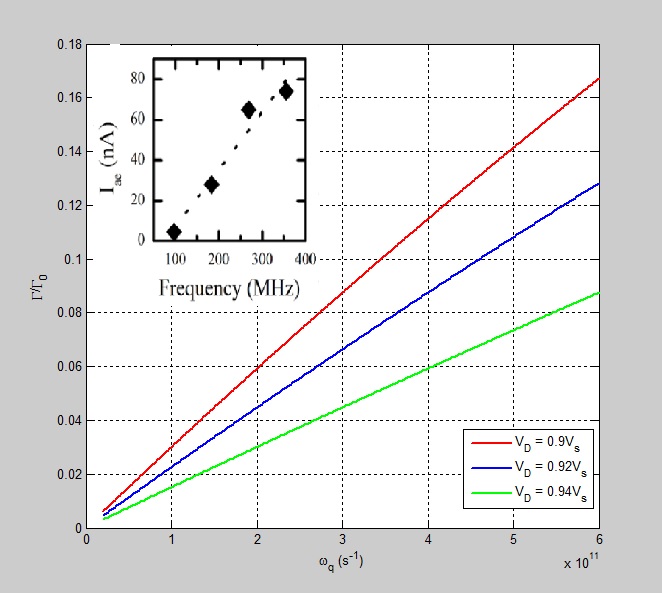}
\caption{ Dependence of $\Gamma/\Gamma_0$ on $\omega_q$ insert is the experimental
verification of Acoustoelectric current versus acoustic phonon frequency ~\cite{37}}
\end{centering}
\end{figure}
\begin{figure}
\includegraphics[width =7.0cm]{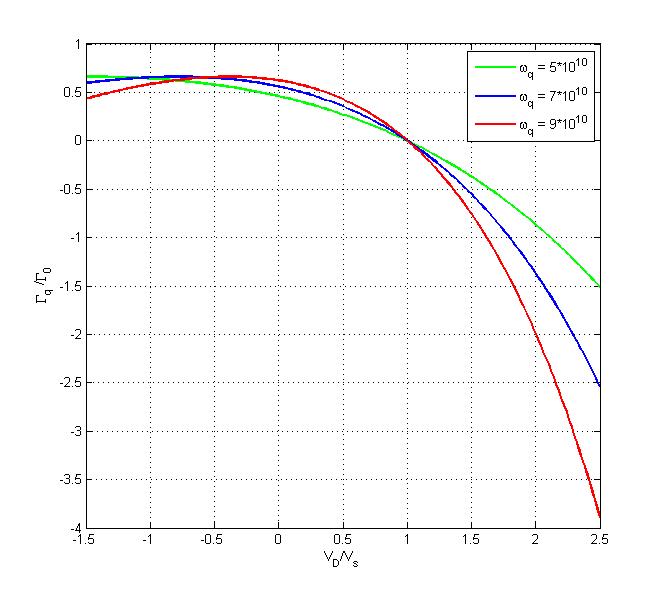}
\includegraphics[width =7.0cm]{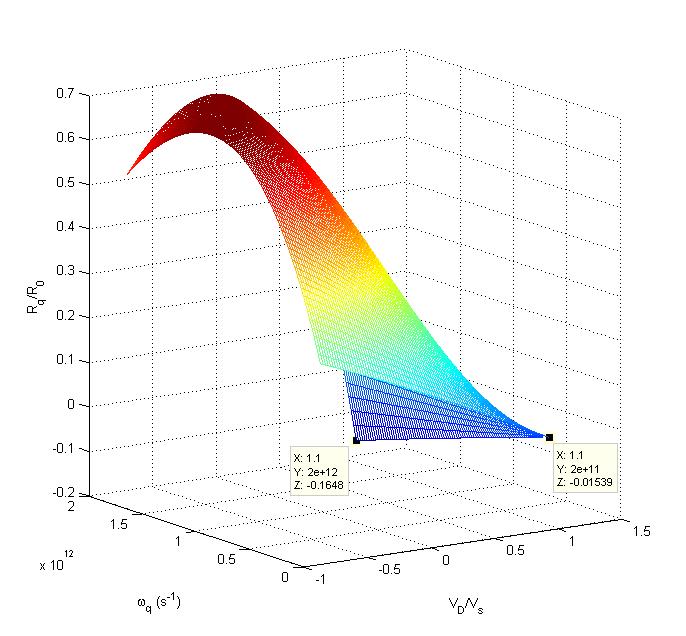}
\caption{  (a) Dependence of $\Gamma/\Gamma_0$ on $V_D\over V_s$ for varying  $\omega_q$ (left) 
(b) $3D$ representation of $\Gamma/\Gamma_0$ on $V_D\over V_s$ and $\omega_q$ at $0.2THz$ (right)}
\end{figure}
\begin{figure}
\includegraphics[width =7.0cm]{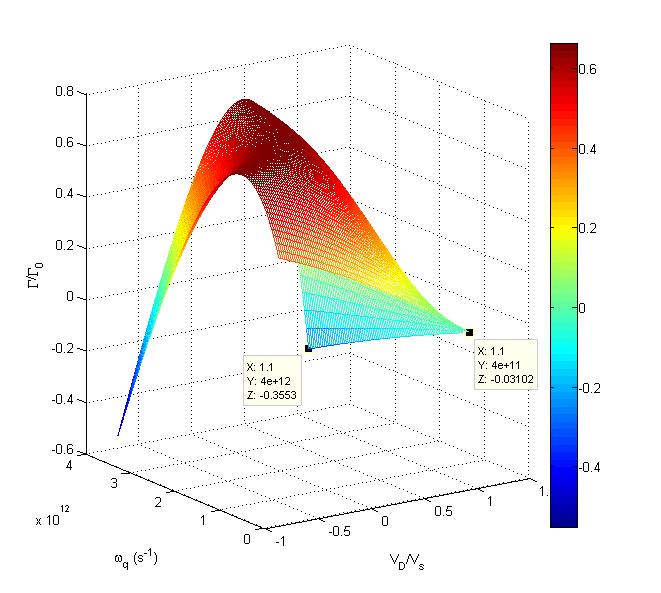}
\includegraphics[width =7.0cm]{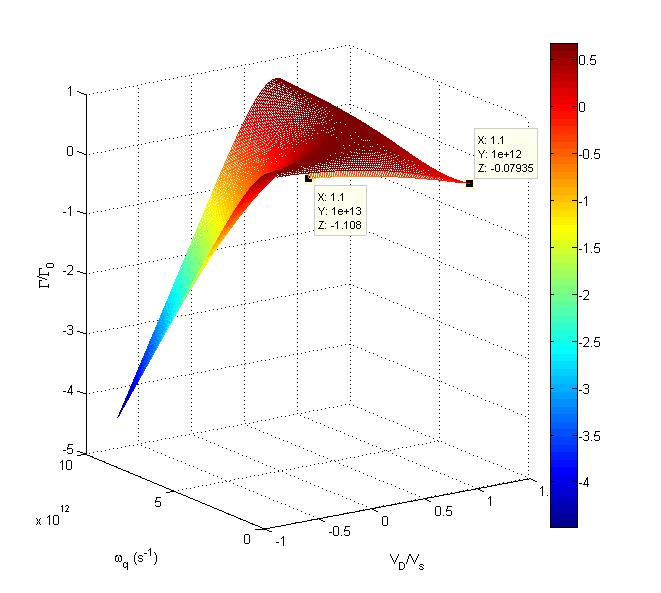}
\caption{ $3D$ representation of $\Gamma/\Gamma_0$ on $V_D\over V_s$ and $\omega_q$ at (a) $0.4THz$ (left)
and at $1THz$(right)}
\end{figure}
It is interesting to note that, acoustic-phonon frequencies above $10THz$ can be attained.  In 
Figure $4$, at $V_D = 1.1V_s$, gave ${\Gamma\over\Gamma_0} = -3.17$ which was obtained  at $\omega_q = 20THz$

\begin{figure}
\begin{centering}
\includegraphics[width = 9.0cm]{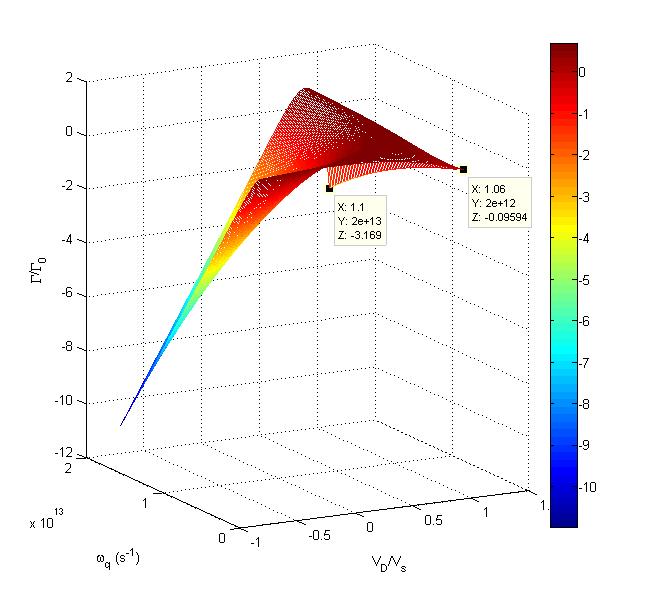}
\caption{ A graph of $\Gamma/\Gamma_0$ on $V_D\over V_s$ and $\omega_q$ at $2THz$}
\end{centering}
\end{figure}
For graphene, with $V_D = 1.1V_s$, the field $E$ can be calculated since $E = V_D/ \mu$. The electron mobility $\mu$ in graphene
given as $2.0\times 10^4 cm^2/Vs$, $V_s = 2.1\times 10^5cm/s$ gives $ E =  11.5V/cm$. 

\section*{Conclusion}
The generation of hypersound amplification (absorption) of acoustic - phonons in  Graphene
is studied.  For ${V_D\over V_s} > 1$, the hypersound amplification obtained is similar to that of Nunes and Fonseca
but  for ${V_D\over V_s} < 1$, an absorption is obtained which could lead to Acoustoelectric Effect in Graphene.
The absorption obtained qualitatively agreed with an experimentally 
obtained acoustoelectric current in a graphene via the Weinrich relation. 
For a drift velocity of $V_D = 1.1V_s$, a field of $E = 11.5V/cm$ was calculated. At frequency of $0.2THz$, 
an amplification of $\Gamma/\Gamma_0 = -3.17$  is attained.  From this work, the hypersound 
studies in graphene offers a much better source of higher phonon 
frequencies  than the homogenous semiconductors which permit the use of graphene as hypersound 
phonon laser (SASER).

\renewcommand\refname{Bibliography}


\begin{thebibliography}{99} 
\bibitem{1}
Novoselov KS, Geim AK, Morozov SV, Jiang D, Zhang Y, Dubonos SV,
Grigorieva IV, Firsov AA: Electric field effect in atomically thin carbon
films. Sci 2004, 306:666.
\bibitem{2}
Castro Neto AH, Guinea F, Peres NM, Novoselov KS, Geim AK, Rev, Mod:
The electronic properties of graphene. Phys 2009, 81:109.
\bibitem{3}
L. Chico et al, Phys. Rev. Lett. 76, 971 (1996).
\bibitem{4}
P. L. McEuen, M. S. Fuhrer, H. Park, IEEE Trans. Nanotechnol. 1, 78 (2002).
\bibitem{5}
Booth, T. J., Blake, P., Nair, R. R., Jiang, D., Hill, E. W., Bangert, U.,
Bloch, A., Gass, M., Novoselov, K. S., Katsnelson, M. I., and Geim, A. K.,
Nano Lett. 8, 2442 (2008).
\bibitem{6}
Willett,R. L., West, K. W.  and Pfeiffer, L. N. , Phys. Rev. Lett. 78, 4478 (1997).
\bibitem{7}
 Shapere, A., and Wilczek, F., (eds.), Geometrical Phases in Physics, World Scientific, 
Singapore, (1989)
\bibitem{8}
 Novoselov, K. S., Geim, A. K.,  Morozov, S. V., Jiang, D.,   Katsnelson, M. I.,   Grigorieva, I. V.,
 Dubonos,  S. V., and  Firsov, A. A., Nature 438, 197 (2005).
\bibitem{9}
Lin, Y. M., Dimitrakopoulos, C., Jenkins, K. A., Farmer, D. B., Chiu, H. Y., Grill, A., Avouris, P.,
Science 327, 662 (2010). 
\bibitem{10}
Xu, W., Gonp, Y. P., Liu, L. W., Qin, H., and Shi, Y. L., Nanoscale Res. Lett. 6, 250 (2011).
\bibitem{11}
Ryzhii, V., Satou, A., and Otsuji, T., J. Appl. Phys. 101, 024509 (2007). 
\bibitem{12} 
Komirenko, S.V. , Kim, K. W. ,Demidenko, A. A. ,Kochelap, V. A., Strocio, M. A., ``Cerenkov generation of high-frequency confined 
acoustic phonons in quantum wells'' , arXiv:cond-mat/9911381v1 23-11-1999. 
\bibitem{13}
Tolpygo K B and Uritskii Z I 1956 Zh. Eksp. Teor. Fiz. 30 929
\bibitem{14}
Weinreich G. 1956 Phys. Rev. 104 321.
\bibitem{15}
Pomerantz M., (1964) Phys. Rev. Lett. 13 308.
\bibitem{16}
Lee, Y. C. and Tzoar N., Phys. Rev. 178, 3, 1969.
\bibitem{17}
 Mensah,S. Y., Allotey, F. K. A.,  Mensah,N. G.,  Elloh,V. W., Amplification of acoustic phonons in a degenerate semiconductor
superlattice. Physica E, Vol. 19(3) ,2003
\bibitem{18}
 S. Y. Mensah, F.K.A. Allotey and S.K. Adjepong, J. Phys.: Condens. Matter 6, (1994) 6793
\bibitem{19}
S. Y. Mensah, F.K.A. Allotey and N.G. Mensah, J. Phys.: Condens. Matter 12, (2000) 5225
\bibitem{20}
S. Y. Mensah, F.K.A. Allotey and N.G. Mensah, H. Akrobotu, G. Nkrumah, Superlattice 
and Microstructure 37 (2005) 87 – 97
\bibitem{21}
Bau, N. Q. and Hieu, N. V., The Influence of the Electromagnetic Wave on the Quantum Acoustomagnetoelectric Field
in a Quantum Well with a Parabolic Potential, PIERS Proceedings, Guangzhou, China, Aug. 25-28, 2014.
\bibitem{22}
Kasala Suresha,  S.S. Kubakaddi, B.G. Mulimani,  Shyi Long Lee, Acoustic wave ampli?cation in 
one-dimensional quantum well wires, Physica E, 33, 50-56 (2006).
\bibitem{23}
Dompreh, K. A.,  Mensah, S. Y.,  Abukari, S. S.,  Sam, F., Mensah, N. G., 
Amplification of Acoustic Waves in Graphene Nanoribbon in the Presence of External Electric and Magnetic Field, 
arXiv.1410.8064v3, (2014).
\bibitem{24}
Zhao, X. F., Zhang, J., Chen, S. M. and Xu, W., ``Cerenkov acoustic-phonon emission generated electrically
from a polar semiconductor", J. Applied Physics 105, (2009).
\bibitem{25}
Heinonen, O.,  Taylor, P. L.,  and Girvin, S. M., Phys. Rev. B 30, 3016 (1984).
\bibitem{26}
Wahlstrand, J. K., Sterens, T. E., Kuhl, J. and Merlin, R., Physica B 316-317, 55 (2002).
\bibitem{27}
Sun, C. K., Chern, G. W., Lin, K. H. and Huang, Y. K., Chin. J. Phys. Taipei 41, 643 (2003).
\bibitem{28}
Beardsley, R. P., Akimov, A. V., Henini, M. and Kent, A. J., Phys. Rev. Lett. 104, 085501 (2010).
\bibitem{29}
Grudinin, I. S., Lee, H., Painter, O., and Vahala K. J., Phys. Rev. Lett. 104, 083901 (2010). 
\bibitem{30}
Zhao, C. X., Xu, W., and Peeters, F. M.,Cerenkov emission of terahertz acoustic -phonons from graphene,
Applied Phys. Lett. 102, 222101 (2013).
\bibitem{31}
Insepov, Z.,  Emelin, E.,  Kononenko, O.,  Roshchupkin, D.V., Tnyshtykbayev, K.B.,  Baigarin, K.A.,
Surface Acoustic Wave Amplification by DC-Voltage Supplied to Graphene Film, 	arXiv:1410.4712 (2014).
\bibitem{32}
Nunes O. A. C. and Fonseca A. L. A., Amplification of hippersound in graphene under external direct
current electric field, Journal of Applied Physics 112, 043707 (2012).
\bibitem{33}
Xia, F.,   Farmer, D. B.,  Lin,Y.-M. and  Avouris, P. Nano Lett. 10, 715 (2010).
\bibitem{34}
Echtermeyer, T. J.,  Lemme, M. C., Bolten, J.,   Baus, M.,  Ramsteiner, M. and
 Kurz, H. Eur. Phys. J. Spec. Top. 148, 19 (2007).
\bibitem{35}
 Barreiro, A.,  Lazzeri, M., Moser,  J., Mauri, F. and Bachtold, A.,  Phys. Rev.
Lett. 103, 076601 (2009).
\bibitem{36}
Weinreich G. 1956 Phys. Rev. 104 321.
\bibitem{37}
 Bandhu, L.,Lawton, L. M.  and Nash G. R. ``Macroscopic acoustoelectric charge transport in graphene",
App. Phys. Lett. 103, 133101 (2013).





\end{thebibliography}
\end{document}